\documentstyle[preprint,aps,epsfig]{revtex}
\tighten
\def\bea{\begin{eqnarray}}
\def\beann{\begin{eqnarray*}}
\def\beq{\begin{equation}}
\def\eea{\end{eqnarray}}
\def\eeann{\end{eqnarray*}}
\def\eeq{\end{equation}}
\def\nn{\nonumber}

\newcommand{\bb}{\bbox{b}}

\newcommand{\br}{\bbox{r}}

\newcommand{\bcdot}{\bbox{\cdot}}

\newcommand{\bk}{\bbox{k}}
\newcommand{\bp}{\bbox{p}}

\newcommand{\bG}{\bbox{G}}

\newcommand{\bgamma}{\bbox{\gamma}}

\newcommand{\bsigma}{\bbox{\sigma}}

\newcommand{\btau}{\bbox{\tau}}
\newcommand{\bpi}{\bbox{\pi}}

\begin{document}
\headheight1.2cm
\headsep1.2cm
\baselineskip=20pt plus 1pt minus 1pt
%\tolerance=1500
%
\preprint{
\vbox{
\hbox{IFT-P.069/98, ADP-98-66/T333}
}}
%%%%%%%%%%%%%%%%%%%%%%%%%%%%%
\draft
\title{Fock terms in the quark-meson coupling model}
\author{
G. Krein$^{1,2}$, A.W. Thomas$^1$ and K. Tsushima$^1$ \\
{\small $^1$Department of Physics and Mathematical Physics and Special 
Research Center for} \\
{\small the Subatomic Structure of Matter, University of Adelaide, SA 5005, 
Australia} \\
{\small  $^2$ Instituto de F\'{\i}sica Te\'{o}rica, Universidade Estadual 
Paulista}\\
{\small Rua Pamplona, 145 - 01405-900 S\~{a}o Paulo, SP, Brazil}
}
\maketitle
\begin{abstract}
The mean field description of nuclear matter in the quark-meson coupling 
model is improved by the inclusion of exchange contributions (Fock terms). 
The inclusion of Fock terms allows us to explore the momentum dependence of 
meson-nucleon vertices and the role of pionic degrees of freedom in matter.
It~is~found that the Fock terms maintain the previous predictions of the
model for the in-medium properties of the nucleon and for the nuclear
incompressibility. The Fock terms significantly increase the absolute 
values of the single-particle, four-component scalar and vector potentials, 
a feature that is relevant for the spin-orbit splitting in finite nuclei.
\end{abstract}
\vspace{2.5cm}
\noindent{PACS NUMBERS: 21.65.+f, 24.85.+p, 24.10.Jv, 12.39.-x}

\vspace{1.0cm}

\newpage 
\hspace*{-\parindent}{\bf 1. Introduction}\hspace*{\parindent}

The quark-meson coupling (QMC) model~\cite{QMC} provides a simple extension 
of relativistic many-body models based on point-like hadrons, such as 
quantum hadrodynamics (QHD)~\cite{QHD}, to include the explicit quark 
structure of the hadrons. Although both QMC and QHD and their variations
share many 
common features, there exist prominent differences. Perhaps the most 
striking one is the mechanism through which the effective masses of the light 
vector mesons ($\rho$, $\omega$ and $\phi$) decrease 
in medium. While in QMC 
the decrease results from an increase in the lower component of the Dirac 
spinor of the quarks in the mesons~\cite{vm-QMC}, in QHD the decrease is 
driven by the vacuum polarization in medium~\cite{vm-QHD}. Modern versions 
of QHD, based on the ideas of effective field theory incorporate vacuum 
polarization effects implicitly through phenomenological effective 
couplings~\cite{QHD-EFT}. However, one can obtain either an increase or
a decrease of the masses 
of the vector mesons when different sets of effective 
couplings are used -- even though they fit ground-state 
observables of nuclei equally well.

The QMC model has been applied to a great variety of problems in nuclear 
physics using the mean field approximation~\cite{QMC-list}. In this paper
we extend the model to include the exchange, or Fock terms, which are 
required by the Pauli exclusion principle. 
As in previous applications of QMC, 
we consider nuclear matter as a system of non-overlapping bags and 
effects due to quark-exchange between different bags~\cite{QMC-qexch} are 
neglected. Thus the Pauli principle is enforced at the nucleon level.
It is important to consider Fock terms not only as a matter of 
principle, but also because it is only through them that the momentum 
dependence of the meson-nucleon vertices can be explored. 
As we know, for example, from the work of Bouyssy et al.~\cite{franc} within 
QHD, the momentum dependence of the exchange contributions from 
the isovector mesons ($\pi$ and $\rho$) is essential to explain the 
magnitude and the systematic behaviour with N and Z of the 
spin-orbit splittings in finite nuclei. 

In the next section we formulate the QMC model in the language of an effective
meson-nucleon Hamiltonian following the ideas of the Cloudy Bag 
Model~(CBM)~\cite{CBM}. In Section~3 we obtain the expressions for the nuclear 
matter energy density and of the single-particle Fock potential. The numerical
results are presented and discussed in Section~4. In Section~5 we present our 
conclusions and discuss perspectives for future calculations.

\newpage
\hspace*{-\parindent}{\bf 2. Effective meson-nucleon Hamiltonian }
\hspace*{\parindent} 

The Lagrangian density of the model is
\bea
{\cal L}_{\text{QMC}} &=& {\cal L}_{\text{MIT}} 
+ \frac{1}{2}\left(\partial_{\mu}\sigma\partial^{\mu}\sigma 
- m^2_{\sigma}\sigma^2\right) - \frac{1}{4}F_{\mu\nu}F^{\mu\nu}
+ \frac{1}{2} m^2_{\omega}\,\omega_{\mu}\omega^{\mu} \nn\\
&+& \frac{1}{2}\left(\partial_{\mu}\bpi\bcdot\partial^{\mu}\bpi 
- m^2_\pi \bpi^2 \right)
- \frac{1}{4}\bG_{\mu\nu}\bcdot\bG^{\mu\nu} + \frac{1}{2}m^2_{\rho}\,
\bb_{\mu}\bcdot\bb^{\mu} \nn\\
&+&\bar\psi\left( g^q_{\sigma}\sigma  
- g^q_{\omega}\gamma_{\mu}\omega^{\mu} 
- \frac{1}{2f_\pi}\gamma_{\mu}\gamma_5\btau\bcdot\partial_{\mu}\bpi
- g^q_{\rho}\frac{1}{2}\gamma_{\mu}\btau\bcdot\bb^{\mu} \right)\psi \,
\theta_V 
\label{Lag}
\eea
where ${\cal L}_{\text{MIT}}$ is the Lagrangian density of the MIT Bag 
Model,
\beq
{\cal L}_{\text{MIT}} = \left[\bar\psi\left(i\gamma_{\mu}\partial^{\mu}
-m_q\right)\psi-B\right]\theta_V -\frac{1}{2}\bar\psi\psi\,\delta_S ,
\label{LagMIT}
\eeq
$ \theta_V$ is one inside the bag and zero outside, $\delta_S$ is 
a delta function on the bag surface,
$F_{\mu\nu}=\partial_{\mu}\omega_{\nu}-\partial_{\nu}\omega_{\mu}$, and
$\bG_{\mu\nu}=\partial_{\mu}\bb_{\nu}-\partial_{\nu}\bb_{\mu}$.
The $\sigma$ and $\omega^{\mu}$ field operators have constant expectation 
values in symmetric nuclear matter, while $\bb^\mu$ field has a nonzero
expectation value only in asymmetric nuclear matter, and the $\bpi$ field has 
zero expectation value in both symmetric and asymmetric matter, because of 
parity considerations. We then separate from the meson field operators their 
mean field values and treat the fluctuations in time-ordered perturbation 
theory. More specifically, we begin by solving the single-nucleon problem in 
the presence of the constant mean-fields. Then we project the quark-meson 
Hamiltonian, obtained from the fluctuating meson fields coupled to the quarks,
onto the space of the single-nucleon states in the presence of the constant 
mean meson fields. 

The Hamiltonian resulting from this procedure is similar to the usual CBM 
Hamiltonian, with the difference that the effective meson-nucleon 
vertices in the present case are evaluated with density dependent wave 
functions. In this approach exchange effects from quarks in different 
nucleons are not taken into account. We also note that our calculation is not
fully self-consistent, in the sense that the Fock terms are calculated 
perturbatively using Hartree self-energies. Our approach is similar to Chin's 
in QHD~\cite{Chin}. There, the difference between the perturbative and full, 
self-consistent results was not very large, because the Hartree terms 
provide by far the most important contributions. Since the Hartree terms also
dominate in QMC, we expect that such a perturbative approach should
be sufficiently reliable for a first estimate of the Fock terms in this case,
and reserve for a future work a more complete, self-consistent calculation.

Let $\sigma_0$, $\omega_0$ and $b_0$ denote the meson  mean fields. Let 
$B^{\dag}_{\lambda}(\bp)$ and $B_{\lambda}(\bp)$ denote the creation and
annihilation operators for single-nucleon states with spin-isospin $\lambda$
and momentum $\bp$ in the presence of the mean fields. That is, the operator 
$B^{\dag}_{\lambda}(\bp)$ creates the state
\beq
|\bp,\lambda\rangle = B^{\dag}_{\lambda}(\bp)|0\rangle,
\label{single}
\eeq
which has energy 
\beq
\varepsilon(\bp,\lambda) = E^*(\bp) + 3 g^q_{\omega}\,\omega_0
+ \frac{1}{2}g^q_{\rho}\,\langle\lambda|\tau_3|\lambda \rangle\,b_0 \, ,
\label{epsilon}
\eeq
with $E^*(\bp) = \sqrt{\bp^2 + M^{*2}}$.
In the present paper we adopt the nonrelativistic normalization, 
\beq
\langle\bp',\lambda'|\bp,\lambda\rangle = \delta(\bp'-\bp)\,
\delta_{\lambda'\lambda} .
\label{normstate}
\eeq
The effective nucleon mass, $M^*$, is obtained by solving the MIT bag 
equations in the presence of the mean fields. It is given explicitly 
by~\cite{GSRT} 
\beq
M^* = \frac{3 \Omega - z_0}{R^*} + \frac{4}{3}{\pi}BR^{*3} ,
\label{M*}
\eeq
where we have taken equal up and down quark masses, $z_0$ is the parameter 
that takes into account zero-point and c.m. motion and
\beq
\Omega = \sqrt{x^{*2} + (R^* m^*_q)^2} ,
\label{Omega}
\eeq
with 
\beq
m^*_q = m_q - g^q_{\sigma} \sigma_0. 
\label{m*}
\eeq
Here, $x^*$ is the solution of the transcendental equation
resulting from the linear (confining) boundary condition at the bag surface:
\beq
j_0(x^*)= \sqrt{ \frac{\Omega - R^* m^*_q}{\Omega + R^* m^*_q}}\; j_1(x^*) ,
\label{trans}
\eeq
where $j_0$ and and $j_1$ are spherical Bessel functions. The in-medium 
radius, $R^*$, is obtained by minimizing $M^*$ with respect to $R^*$. 

The next step is to project the quark-meson Hamiltonian, obtained from 
the Lagrangian density given in Eq.~(\ref{Lag}), onto the single-nucleon 
states of Eq.~(\ref{single}). The resulting effective meson-nucleon 
Hamiltonian can be written as
\beq
H_{\text{eff}} = H_{\text{MMF}} + H_0 + W,
\label{Heff}
\eeq
where $H_{\text{MMF}}$ is the Hamiltonian of the meson mean fields,
\beq
H_{\text{MMF}} = \frac{1}{2}m^2_\sigma \sigma^2_0 
- \frac{1}{2} m^2_\omega \,\omega^2_0 
- \frac{1}{2} m^2_\rho \,b^2_0 ,
\label{HMF}
\eeq
$H_0$ is the sum of the single-nucleon and single-meson Hamiltonians, and
$W$ is the meson-nucleon interaction. Specifically, $H_0$ is the sum
\beq
H_0 = \sum_{\lambda} \int d\bp \, 
\varepsilon(\bp,\lambda)\,B^{\dag}_{\lambda}(\bp)\,B_{\lambda}(\bp) 
+ \sum_{j=\sigma,\omega,\pi,\rho}\int d\bk \,\omega_{j}(\bk)\,
a^{\dag}_j(\bk)\,a_j(\bk) ,
\label{H0}
\eeq
where $\varepsilon(\bp,\lambda)$ is given in Eq.~(\ref{epsilon}),
$\omega_{j}(\bk) = (\bk^2 + m^2_j)^{1/2}$, and $a^{\dag}_j(\bk)$ and 
$a_j(\bk)$ are the meson creation and annihilation operators. 
The meson-nucleon interaction, $W$, can be written as
\beq
W = \sum_{j=\sigma,\omega,\pi,\rho}
\int d\bp \,d\bp'\, d\bk \,\delta(\bk-\bp'+\bp)\,
\frac{(2\pi)^{3/2}}{\sqrt{2\omega_j(\bk)}}\sum_{\lambda\lambda'} 
W^j_{\lambda'\lambda}(\bp',\bp)\,
B^{\dag}_{\lambda'}(\bp')\,B_{\lambda}(\bp)\,a_j(\bk) + \text{h.c.} ,
\label{W}
\eeq
where the vertices $W^j_{\lambda'\lambda}(\bp',\bp)$ are of the general form
\beq
W^j_{\lambda'\lambda}(\bp',\bp) = \langle\bp'\lambda'|\bar\psi(0)
\Gamma^j\psi(0)|\bp\lambda\rangle,
\label{Wp}
\eeq
with 
\bea
\Gamma^{\sigma} &=& -g^q_{\sigma} , \label{Gammas}\\
\Gamma^{\omega} &=& -g^q_{\omega}\,\not\!\epsilon_{\omega}, 
\label{Gammav} \\
\Gamma^{\pi} &=& i \frac{f^q_{\pi}}{m_{\pi}}\, \gamma^{\mu}(p'-p)_{\mu}
\gamma_5\btau , \label{Gammapi} \\
\Gamma^{\rho} &=& - \frac{1}{2}g^q_{\rho} \not\!\epsilon_{\rho}\,\btau . 
\label{Gammarho}
\eea
Here $k=(E^*(\bp')-E^*(\bp),\bp'-\bp)$, and the polarization vectors 
$\epsilon^{\mu}(\bk)$ satisfy 
\beq
\sum^3_{s=1} \epsilon^{\mu}_{\omega,\rho}(\bk,s) 
\epsilon^{\nu}_{\omega,\rho}(\bk,s)
= - g^{\mu\nu} + \frac{k^{\mu}k^{\nu}}{m^2_{\omega,\rho}} .
\label{completeness}
\eeq

The next step requires evaluation of the matrix elements 
$W^j_{\lambda'\lambda}(\bp',\bp)$. For free-space nucleon states, matrix 
elements of this sort have been calculated in the MIT bag model in a variety 
of approximations. In the present case, the situation 
is more complicated because the states 
$|\bp,\lambda\rangle$ have their mass-shell modified in medium. 
Not only is the 
nucleon mass modified from its free-space value, $M$, to $M^*$, but the 
relation between energy and momentum is also 
modified by the mean vector fields, 
as can be seen in Eq.~(\ref{epsilon}). These modifications are, of course, the 
interesting aspect of the QMC model as they will introduce a density 
dependence into the form factors. In order 
not to complicate the problem too much, 
in this initial study we adopt a simple approach. First, we parametrize the 
various matrix elements $W^j_{\lambda'\lambda}(\bp',\bp)$ as in free space, 
namely
\beq
W^{j}_{\lambda'\lambda}(\bp',\bp) = \bar u(\bp',\lambda')\,O^{j}(k)\,
u(\bp,\lambda) ,
\label{WandO}
\eeq
with  
\bea
O^{\sigma}(k) &=& - g^q_{\sigma}\, F_s(k^2)\, \label{Os} \\
O^{\omega}(k) &=& - g^q_{\omega}\, 
\epsilon^{\mu}_{\omega} \,\left[\gamma_{\mu} F^{(\omega)}_1(k^2) + 
\frac{i\sigma_{\mu\nu}k^{\nu}}
{2M^*} F^{(\omega)}_2 (k^2)\right] , \label{Ow} \\
O^{\pi}(k) &=& i \frac{1}{2f_\pi}\,
G_A (k^2) \,\not\! k \gamma_5 \btau , \label{Opi} \\
O^{\rho}(k) &=& - \frac{1}{2}g^q_{\rho}\, 
\epsilon^{\mu}_{\rho}\,\left[\gamma_{\mu} F^{(\rho)}_1(k^2) + 
\frac{i\sigma_{\mu\nu}k^{\nu}}
{2M^*} F^{(\rho)}_2 (k^2)\right] \btau . \label{Orho}
\eea
The Dirac spinors, $u(\bp,\lambda)$, are given by
\beq
u(\bp,\lambda) = \sqrt{\frac{E^*(\bp)+M^*}{2E^*(\bp)}} 
\left(\begin{array}{c}
1 \\
\displaystyle{
\frac{\bsigma\bcdot\bp}{E^*(\bp)+M^*} 
}
\end{array}  \right)\chi_s \xi_{t},
\label{spinor}
\eeq
where $\chi_s$ and $\xi_t$ are the Pauli spinors for spin and isospin,
and $E^*(\bp)=\sqrt{\bp^2+M^{*2}}$. Next, we calculate the various form 
factors in the usual way, using, however, the bag model wave-functions 
modified by the mean fields. In calculating the form factors we ignore 
center-of-mass and recoil corrections, as well as effects due to Lorentz 
contraction when going to the Breit frame~\cite{LichtParg}. Both 
approximations can be improved by using the technique recently developed 
in Ref.~\cite{LTW}, but no qualitative changes with respect to their density
dependence are expected in using the present approximations. We note that in 
Eq.~(\ref{Opi}) we have not included the induced pseudoscalar form factor; 
we intend to investigate its effect on nuclear matter properties when using a
chiral model, such as the CBM. 

Let the single-quark wave functions in the presence of the mean fields be 
written as
\beq
q(\br) = \left(\begin{array}{c}
g(r) \\
\displaystyle{
i\bsigma\bcdot\hat{\br} f(r) }
\end{array}  \right) \phi(\hat{\br}) ,
\label{qspinor}
\eeq
where $\phi(\hat{\br})$ contains the spin-isospin wave functions. In 
calculating the vector meson form factors, we follow the standard
procedure of relating the Dirac form factors $F_1$ and $F_2$ to the Sachs
form factors, $G_E$ and $G_M$, as $F_1 = [G_E + \eta G_M]/(1+\eta)$ and
$F_2= [G_M - G_E]/(1+\eta)$ with $\eta=-k^2/4M^{*2}$. The various form 
factors in Eqs.~(\ref{Os}) -~(\ref{Orho}) are given by:
\bea
F_s(k^2) &=& 3\, \int d\br\, j_0(kr) \left[g^2(r)-f^2(r)\right] , \label{Fs}\\
G^{\omega}_E(k^2) &=& 3\, \int d\br\, j_0(kr) \left[g^2(r)+f^2(r)\right] 
\equiv 3\,G_E(k^2) ,
\label{GEome}\\
G^{\omega}_M(k^2) &=& 2 M^* \int d\br\, \frac{j_1(kr)}{k} 
\left[2g(r)f(r)\right] \equiv G_M(k^2) , 
\label{GMome}\\
G_A(k^2) &=& \frac{5}{3}\,\int d\br\, j_0(kr) 
\left[g^2(r)-\frac{1}{3}f^2(r)\right] , \label{GA}\\ 
G^{\rho}_E(k^2) &=& \frac{1}{3}\,G^{\omega}_E(k^2) = G_E(k^2) ,
\label{GErho}\\
G^{\rho}_M(k^2) &=& \frac{5}{3}\,G^{\omega}_M(k^2) = \frac{5}{3}\, G_M(k^2) .
\label{GMrho}
\eea
Note that $G_E$ and $G_M$ (without meson indices) are the usual,  
electromagnetic Sachs form factors.

\vspace{0.5cm}
\hspace*{-\parindent}{\bf 3. Nuclear matter energy and the Fock potential}.
\hspace*{\parindent} 

The energy density of symmetric nuclear matter (in 
which case the mean field $b_0$ does 
not contribute) is given by the sum of the mean-field energies and the
Fock energy, 
\beq
{\cal E}_{NM} = 4 \int \frac{d\bp}{(2\pi)^3}\, \theta(k_F-|\bp|) \,
E^*(\bp)\, 
+ 3\,g^q_{\omega}\omega_0\,\rho_N +  \frac{1}{2}m^2_{\sigma}\sigma^2_0
- \frac{1}{2} m^2_{\omega}\omega^2_0 
+ \sum_{j=\sigma,\omega,\pi,\rho}{\cal E}^{j}_{\text{Fock}} ,
\label{EMF} 
\eeq
where $\rho_N$ is the nucleon density and the Fock energies 
${\cal E}^{j}_{\text{Fock}}, j =\sigma, \omega, \pi, \rho$ are the 
second-order exchange contributions
\beq
{\cal E}^{j}_{\text{Fock}} = \frac{1}{2} \int \frac{d\bp}{(2\pi)^3} 
\, \theta(k_F-|\bp|)
\int \frac{d\bp'}{(2\pi)^3}\, \theta(k_F-|\bp'|)\, 
\frac{\sum_{\lambda\lambda'}|W^j_{\lambda\lambda'}(\bp,\bp')|^2}
{(\bp'-\bp)^2+m^2_j} .
\label{EFock} 
\eeq
 
Using Eqs.~(\ref{Os})-(\ref{Orho}) in the expression for the Fock energy 
density, Eq.~(\ref{EFock}), we can rewrite it as
\beq
{\cal E}^{j}_{\text{Fock}} = \frac{1}{2}\int \frac{d\bp}{(2\pi)^3}
\, \theta(k_F-|\bp|) \sum_{\lambda} \bar u(\bp,\lambda)\,
U^j_{\text{Fock}}(\bp)u(\bp,\lambda)
\eeq
with the ``Fock potential", $U^j_{\text{Fock}}(\bp)$, or the exchange nucleon 
self-energy given by
\beq
U^j_{\text{Fock}} = \int \frac{d\bp}{(2\pi)^3}\, \theta(k_F-|\bp|) 
O^j(-k)\,\sum_{\lambda'}u(\bp',\lambda')\bar u(\bp',\lambda')\,
O^{j}(k) ,
\eeq
where the four-momentum $k$ is given as before. The potential 
$U^j_{\text{Fock}}(\bp)$ for each meson has the Dirac structure,
\beq
U^j_{\text{Fock}}(\bp) = U^j_s(\bp) + \gamma^0 \,U^j_0 (\bp) 
+ \bgamma\bcdot\hat{\bp}\,U^j_{v} .
\label{DiracU}
\eeq
The explicit forms of the potentials are easily obtained by making use of 
Eqs.~(\ref{Os})-(\ref{Orho}) and will be given in a separate publication.

The final step is to determine the scalar 
mean field, $\sigma_0$, which satisfies
the self-consistency equation 
\beq
\sigma_0 = \frac{g_{\sigma}}{m^2_\sigma}\,C(\sigma_0) \, 4 
\int \frac{d\bp}{(2\pi)^3}\, \theta(k_F-|\bp|) \frac{M^*}{E^*(\bp)} + 
\frac{1}{m^2_{\sigma}}\, \frac{\partial} {\partial\sigma_0}
\sum_{j=\sigma,\omega,\pi,\rho}{\cal E}^{j}_{\text{Fock}} ,
\label{sigma0}
\eeq
where
\beq
g_{\sigma} = 3 g^q_{\sigma} S(0),\hspace{1.5cm}
C(\sigma_0) = \frac{S(\sigma_0)}{S(0)},
\label{gC}
\eeq
with
\beq
S(\sigma_0)= \frac{ \Omega/2 + m^*_q R^* (\Omega-1) }
{\Omega(\Omega-1) + m^*_q R^*/2 } .
\label{S}
\eeq
The mean vector field, $\omega_0$, is given in terms of the
baryon density, as usual.

\vspace{0.5cm}
\hspace*{-\parindent}{\bf 4. Numerical results}
\hspace*{\parindent} 

Before presenting the numerical results, we discuss some technical 
points. As is well known, the sharp surface of the bag induces 
oscillations in the form factors at large momenta. The process of
removing spurious centre of mass motion and projecting onto a definite
momentum tends to smooth this, but it is extremely time consuming.
In order to avoid this time consuming calculation in this first
investigation, we have chosen 
to parametrize the quark wave functions in terms of smooth functions -- 
following Ref.~\cite{Duck} we use a gaussian form with 
two parameters ($R_0$ and $\beta$). These are adjusted 
to reproduce the r.m.s. radius and the 
quark scalar density of the nucleon calculated with the original MIT bag wave 
functions. Note that $R_0$ and $\beta$ are density dependent. 

We also observe that when c.m. corrections and the meson cloud are 
neglected in the calculation of form factors, some physical quantities such 
as $g_A$ and the magnetic moments are not well reproduced within the model.
For example, $g_A = G_A(0) \sim 1.09$ for the ``bare'' bag without c.m. 
corrections, whereas the experimental value is $g_A=$~1.267~\cite{pdata}. 
The magnetic moments are 
given in the tensor couplings of the $\omega$ and $\rho$,
\bea
\kappa^S &=& \mu_p + \mu_n -1 = \frac{g^q_{\omega}F^{\omega}_2(0)}
{g^q_{\omega}F^{\omega}_1(0)} 
= -1 + \frac{1}{3}\,G_M(0) , \nn\\
\kappa^V &=& \mu_p - \mu_n -1 = 
\frac{g^q_{\rho}F^{\rho}_2(0)}{g^q_{\rho} F^{\rho}_1(0)}
= -1 + \frac{5}{3}\,G_M(0) .
\label{kappas}
\eea
Pionic corrections are known~\cite{CBM-mm} to be very important for these 
quantities. For the nuclear matter calculation, the main effect of neglecting 
these is to underestimate the tensor coupling of the $\rho$ to the nucleon,
$\kappa^V$ ($\kappa^S$ is a small quantity and does not play a big role for
the binding energy of nuclear matter). Therefore, in order to obtain a more
realistic estimate of the contribution of the tensor coupling of the $\rho$ 
we adjust $\kappa^V=\mu_p - \mu_n -1$ to its experimental value. We also
adjust $G_A(0)$ to the experimental value of $g_A$. As we said above, the 
important aspect for our calculation is the density dependence of the form 
factors, and we do not expect qualitative changes in a more complete 
calculation.

We begin by investigating the effect of adding 
the Fock energy to the mean field 
energy (Hartree) per nucleon. We use the $\sigma$ and $\omega$ coupling 
constants of the Hartree approximation to calculate the Fock terms, but solve 
Eq.~(\ref{sigma0}) self-consistently to obtain $\sigma_0$ and $M^*$.
The second term in 
Eq.~(\ref{completeness}) does not contribute in the case of $\omega$ coupling 
because of baryon current conservation, and for $\rho$ coupling it gives an
extremely small contribution to the nuclear binding energy and is therefore
neglected. The Hartree coupling constants are fixed by requiring a stable
minimum at $E/A-M=-15.7$~MeV at $\rho_0=0.15$~fm$^{-3}$. For a free bag
radius of 0.8~fm they are 
given~\cite{GSRT} by $g^2_{\sigma}/4\pi=5.40$ and $g^2_{\omega}/4\pi=5.31$, 
where $g_{\sigma}$ is defined in Eq.~(\ref{gC}) and 
$g_{\omega}=3g^q_{\omega}$. We consider first the Fock terms associated
with just the $\sigma$ and $\omega$.
In Figure~1 we show the results. The effect of 
the Fock terms is repulsive and of the order of $5$~MeV. 

In Figure~2 we show the effect of the pion Fock term. The pion gives a 
repulsive contribution of the order of $8$~MeV. Here, as in QHD, we have the 
situation that the effective NN interaction due to pion exchange contains a 
short-range contact interaction. There have been arguments that contact
interactions should not be taken into account in a mean-field calculation, 
since they are suppressed by short-range NN correlations~\cite{franc}. 
In a complete calculation, in which the full ladder diagrams are summed, 
this suppression of short-distance pion exchange is automatic. In a 
Hartree-Fock treatment, there is no unambiguous way to subtract such short 
range pieces from a relativistic interaction. We approach the problem by 
making a static approximation and expanding the nucleon energies in the 
spinors as
\beq
 E^*(\bp) \simeq M^*+\bp^2/2M^*. 
\label{expans}
\eeq
The contact interaction then becomes evident and 
can easily be subtracted. Specifically, the contribution of the pion to the 
energy density of nuclear matter is given by
\bea
{\cal E}^{\pi} &=& \frac{1}{2}\,\times 12\,\int \frac{d\bp}{(2\pi)^3}
\theta(k_F-|\bp|) \int \frac{d\bp'}{(2\pi)^3} \theta(k_F-|\bp'|)
\,\frac{[G_A(p'-p)]^2}{(\bp'-\bp)^2+m^2_{\pi}}\nn \\
&\times & 2M^{*2}\Bigl[E^*(\bp)E^*(\bp')-M^{*2}-\bp\bcdot\bp'\Bigr] .
\label{Epiex}
\eea
Expanding $E^*$ as in Eq.~(\ref{expans}), we obtain
\bea
{\cal E}^{\pi} &\simeq & 6\,\int \frac{d\bp}{(2\pi)^3} \theta(k_F-|\bp|) 
\int \frac{d\bp'}{(2\pi)^3} \theta(k_F-|\bp'|)
\,[G_A(\bp'-\bp)]^2 \, \frac{(\bp'-\bp)^2}{(\bp'-\bp)^2+m^2_{\pi}} \nn\\
&=& 6\,\int \frac{d\bp}{(2\pi)^3} \theta(k_F-|\bp|)
\int \frac{d\bp'}{(2\pi)^3} \theta(k_F-|\bp'|)
\,[G_A(\bp'-\bp)]^2 \, \left[1-\frac{m^2_\pi}{(\bp'-\bp)^2+m^2_{\pi}}\right] .
\label{Epiapp}
\eea
We have recalculated the energy density using the approximate expression,
Eq.~(\ref{Epiapp}), instead of Eq.~(\ref{Epiex}). We found that the 
approximate expression induces an extremely small change in the result, 
indicating that the terms higher than ${\cal O}(p^2)$ neglected in the 
expansion give a negligible contribution to the energy density. The factor 
$1$ in the square brackets in Eq.~(\ref{Epiapp}) is due to the contact 
interaction in the one-pion-exchange. When this factor is
subtracted, the contribution of the pion becomes attractive (dotted line in 
Figure~2), and is of the order of $5$~MeV.

The $\rho$ also gives a repulsive contribution. Because of the large value
of $\kappa^V$, the tensor term gives by far the most important contribution. 
As in the case of the pion, the tensor component contains a contact term. We 
proceed as previously, make a static approximation and subtract the contact 
term. The energy per particle is insensitive to the static approximation. 
For $g_{\rho} (=g^q_{\rho})$ we use the preferred phenomenological value 
$g_{\rho}~=~2~\sqrt{4\pi\times 0.55}$ and readjust the $\sigma$ and $\omega$ 
coupling constants such as to refit $E/A-M=-15.7$~MeV at 
$\rho_0=0.15$fm$^{-3}$. The new values of the coupling constants for a 
quark mass of $m_q=5$~MeV and for three different values for the free bag 
radius, 0.6~fm, 0.8~fm and 1.0~fm are presented in Table~I. The values of 
the ratios of in-medium to free nucleon masses, $M^*/M$, and r.m.s. radii, 
$r^*/r$, at the saturation density are also shown in Table~I. In the last 
two columns of the table we present the values for the nuclear matter 
incompressibility and symmetry energy. To calculate the contribution from 
the Fock terms to the symmetry energy we have used the approximation of 
Bouyssy~et.~al, discussed on page~387 of Ref.~\cite{franc}. Different values
for the quark mass between $0$ and $10$~MeV do not change the qualitative 
results. 

Inspection of Table~I reveals that the new values for the in-medium bag 
parameters have not changed much from their Hartree values~\cite{GSRT}. For
example, for $R=0.8$~fm, we find that $M^*/M=0.83$ and $r^*/r=1.02$. The
corresponding Hartree values are $M^*/M=0.80$ and $r^*/r=1.09$. Also the 
incompressibility of nuclear matter is not changed much by the addition of the
Fock terms. From the Hartree value, $K=280$~MeV, it has changed to 
$K=285$~MeV. The symmetry energy, $a_4$, is almost entirely
determined by the value of $g_{\rho}$, as is well known. Our value for 
$g_{\rho}$, which has {\em not} been adjusted to the symmetry energy, predicts 
$a_4\sim 30$~MeV, independently of the free bag radius. The experimental 
value is close to $33$~MeV. The contribution of the Fock terms is substantial,
of the order of $10$~MeV. In Figure~3 we plot the binding energy as a function 
of $k_F$ corresponding to parameter set (b) in Table~I. For comparison, we 
also present the Hartree solution in the same figure.

Next we discuss the Fock potentials. It is convenient to introduce 
the sums $V_s = -~g_{\sigma}\,\sigma_0 + U_s$ and 
$V_0~=~g_{\omega}\,\omega_0~+~U_0$, which are the combinations that will 
appear in the single-nucleon Dirac equation. The momentum dependence of the 
components $V_s$, $V_0$ and $U_v$ at the saturation density are shown in 
Figure~4. The separate contributions of the different mesons to these are 
shown in Table~II, for the three parameter sets of Table~I.  The addition of 
the Fock terms increases both the $U_s$ and $U_0$ components, and the 
component $U_v$ is very small, for all three sets of parameters. For 
parameter set (b) of Table~I, we find that $|V_s|+|V_0|$ at $|\bp|=k_F$ has 
increased by $67$~MeV, in comparison with
the value in the Hartree approximation.
Such an increase will have important consequences for the spin-orbit 
splittings in finite nuclei. However, a quantitative estimate of such 
effects for finite nuclei requires a dedicated calculation, 
which is complicated by 
the non-locality of the Fock interaction. We leave this for a future study.

We have also investigated the role of the medium dependence of the form
factors on the saturation properties of nuclear matter. As we mentioned
previously, the self consistent change of the internal structure of the
nucleon with the nuclear medium properties is the attractive new aspect of
the QMC model, and it is important to know the role of such effects
on our results. In order to test these effects, we re-evaluated
the Fock terms using the form factors 
of free space bags, and redetermined the $\sigma$ and $\omega$ 
coupling constants so as to obtain $E/A-M=-15.7$~MeV at 
$\rho_0=0.15$~fm$^{-3}$. For $R~=~0.8$~fm, the new values are 
$g^2_{\sigma}/4\pi=$~4.37 and $g^2_{\omega}/4\pi=$~5.24. The ratios 
$M^*/M$ and $r^*/r$ remain almost the same as in Table~I, but the 
incompressibility and the symmetry energy change substantially. The
incompressibility changes from $K~=~285$~MeV to $K~=~264$~MeV, and the
symmetry energy from $a_4~=~30$~MeV  to $a_4~=~36$~MeV. These 
big effects arise from an increase of the Fock term of the $\sigma$ meson. For 
example, to the $6$~MeV change in $a_4$, the $\sigma$ meson alone contributes 
$5$~MeV. The reason for this large effect is simple to understand. The form 
factor $F_s$ in Eq.~(\ref{Fs}) involves the difference of the squares of the 
upper and lower components of the quark wave functions, and this difference is 
strongly density dependent; the difference decreases as the density increases.
This density dependence is at the heart of the saturation mechanism in the QMC
model, and here, this effect is very visible.

\vspace{0.5cm}
\hspace*{-\parindent}{\bf 5. Conclusions and perspectives}
\hspace*{\parindent} 

In this paper we have developed a calculational scheme to introduce Fock
(exchange) terms in the QMC model. Following the ideas of the Cloudy Bag 
Model, we have constructed an effective meson-nucleon Hamiltonian which 
naturally incorporates momentum- and density-dependent vertices. We found 
that the mean-field predictions of the QMC model for the in-medium properties 
of the nucleons are maintained with the incorporation of the Fock terms. 
The nuclear matter incompressibility remains low as in the mean field 
approximation. Without adjusting the coupling constant of the $\rho$ meson, 
the symmetry energy is predicted to be very close to its experimental value. 
The Fock terms increase the value of the four-component, single-nucleon 
scalar and vector potentials, which is a welcome feature for the 
phenomenology of the level splitting in finite nuclei. 

The formulation of the present model can be extended in several ways. 
We intend to incorporate the effects of chiral symmetry and the 
$\Delta(1232)$. As with nucleons in free space, we expect the Cloudy Bag 
Model to provide insight on the role of chiral symmetry on in-medium nucleon 
and nuclear matter properties. The present formulation of the QMC model seems
to be particularly convenient for the implementation of the coupled cluster
method for studying nucleon correlations in matter~\cite{CBM}. We also
intend to perform a fully self-consistent calculation of the Fock terms.
With respect to the spin-orbit splitting in finite nuclei, it might be
interesting to investigate the effect of Fock terms within an extended model
with a density-dependent bag constant. As shown by Jennings and 
collaborators~\cite{Jenn}, when the bag constant is allowed to decrease as a 
function of the density, the description of the spin-orbit splittings is 
improved. Finally, it will be interesting to see whether, given the implicit 
density dependence of masses and coupling constants in QMC, one still needs
non-linear meson-meson couplings for a good phenomenological description of 
finite nuclei.

\vspace{0.3cm}
\hspace*{-\parindent}{\bf Acknowledgments}.\hspace*{\parindent}

This work was supported in part by the Australian Research Council and
the Brazilian agencies CNPq and FAPESP. G.~K. would like to thank the CSSM 
for the financial support and warm hospitality.

%
%TABLE I
%
\begin{table}
\caption{$\sigma$ and $\omega$ coupling constants of the HARTREE+FOCK 
approximation, for three different free bag radii. Also shown are the ratios 
of the in-medium to free-space nucleon masses and r.m.s. 
radii at the saturation density $\rho_0=0.15$~fm$^{-3}$, and the nuclear 
matter incompressibility $K$ and symmetry energy $a_4$. The quarks mass is 
$m_q=5$~MeV in all cases.}
\begin{center}
\begin{tabular}{c|ccccccc} %rrrrrrr}
{}  & $R$ (fm)& $g^2_{\sigma}/4\pi$ & $g^2_{\omega}/4\pi$ & $M^*/M$ 
& $r^*/r$ & $K$ (MeV) & $a_4$ (MeV) \\ 
\tableline
(a) & 0.6     &  4.36  & 5.55  & 0.82 & 1.01 & 289 & 30 \\
(b) & 0.8     &  4.37  & 5.49  & 0.83 & 1.02 & 285 & 30 \\
(c) & 1.0     &  4.23  & 5.03  & 0.84 & 1.02 & 273 & 30  
\end{tabular}
\end{center}
\end{table}

\vspace{3.0cm}
%
%TABLE II
%
\begin{table}
\caption{Contributions of the different mesons to the single-nucleon
potentials for $|\protect\bp|=k_F$ at the saturation density. Columns (a),
(b) and (c) correspond to the parameters in Table~I. The last line shows the 
HARTREE  values. }
\begin{center}
\begin{tabular}{l|rrrr|rrr|rrr}
Potentials (\text{MeV}) & {}    &  {}     & $V_s$  & {}       
                       & {}     & $V_0$   & {}      
                       & {}     & $U_v$   & {}                       \\
\tableline
Parameters     & {}    &  (a)   &         (b)   &  (c)\hspace{5mm}    
                       &  (a)   &         (b)   &  (c)\hspace{5mm}    
                       &  (a)   &         (b)   &  (c)
\\   
\tableline
Direct         & {}    & $-$175   & $-$173    &  $-$164\hspace{5mm} 
                       &   +131   &   +130    &    +119\hspace{5mm}   
                       &   {--}   &    {--}   &    {--}              \\  
Exchange       & $\sigma$  & +14     & +12    &  +10\hspace{5mm}    
                           & +14     & +12    &  +11\hspace{5mm}      
                           &$-$0     & $-$0   & $-$0                 \\
{}             & $\omega$  &$-$44    &$-$39   &$-$31\hspace{5mm} 
                           &  +23    &  +20   &  +16\hspace{5mm}   
                           & $-$2    & $-$2   & $-$2            \\
{}             & $\pi$     &$-$4    & $-$3  & $-$3\hspace{5mm}    
                           &$-$3    & $-$3   & $-$3\hspace{5mm}   
                           &$-$3    & $-$2   & $-$2             \\
{}             & $\rho$    &$-$32   & $-$34  & $-$32\hspace{5mm}    
                           &   +1   &  $-$0  &  $-$1\hspace{5mm}    
                           &   +8   &    +8  &    +7           \\
%
%\\
% 
Total          &   {}      &$-$241    &$-$237     & $-$220\hspace{5mm}   
                           &  +166    &  +159     &   +142\hspace{5mm}  
                           &    +3    &    +4     &     +4          \\
{}             &   {}      &  {}     &  {}    &  {}\hspace{5mm}    
                           &  {}     &  {}    &  {}\hspace{5mm}    
                           &  {}     &  {}    &  {}         \\
Hartree        &   {}      &$-$228    &$-$204     &$-$186 \hspace{5mm}    
                           &  +150    &  +125     &  +108 \hspace{5mm}     
                           &  {--}   &  {--}    &  {--}        
\end{tabular}
\end{center}
\end{table}

\begin{figure}
\epsfxsize=15.cm
\epsfxsize=15.cm
\centerline{\epsfbox{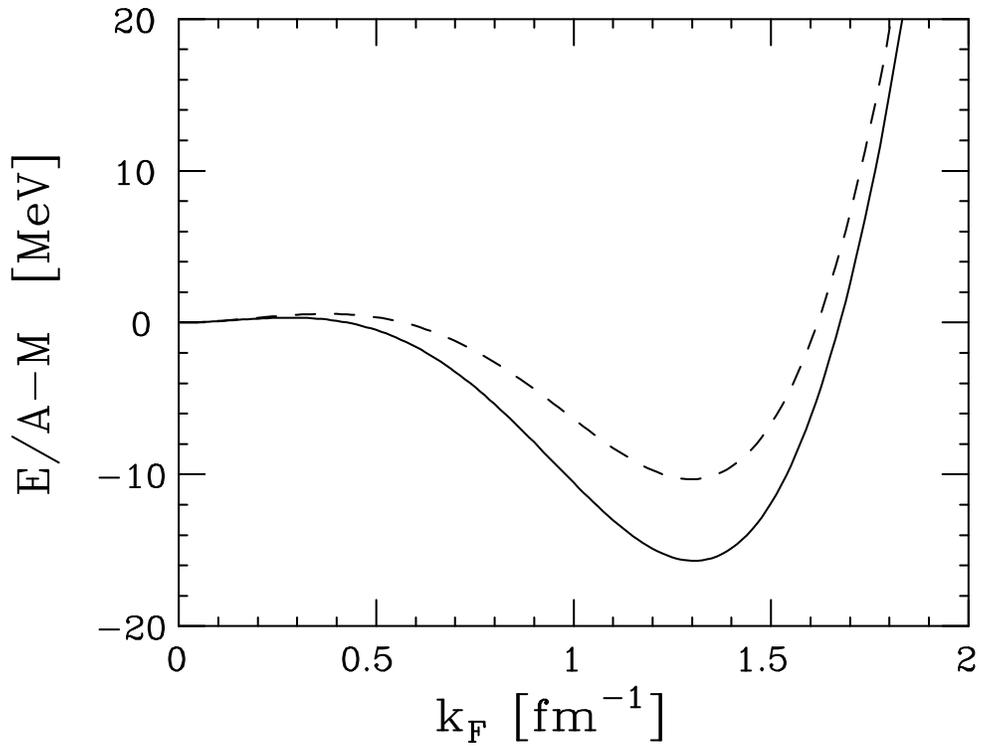}}
\vspace{-3.0cm}
\caption{The energy per particle as a function of the Fermi momentum.
The solid curve is the HARTREE result, and the dashed curve is the 
HARTREE+FOCK result. Here only the $\sigma$ and $\omega$ mesons are 
included.}
\end{figure}

\begin{figure}
\epsfxsize=15.cm
\epsfxsize=15.cm
\centerline{\epsfbox{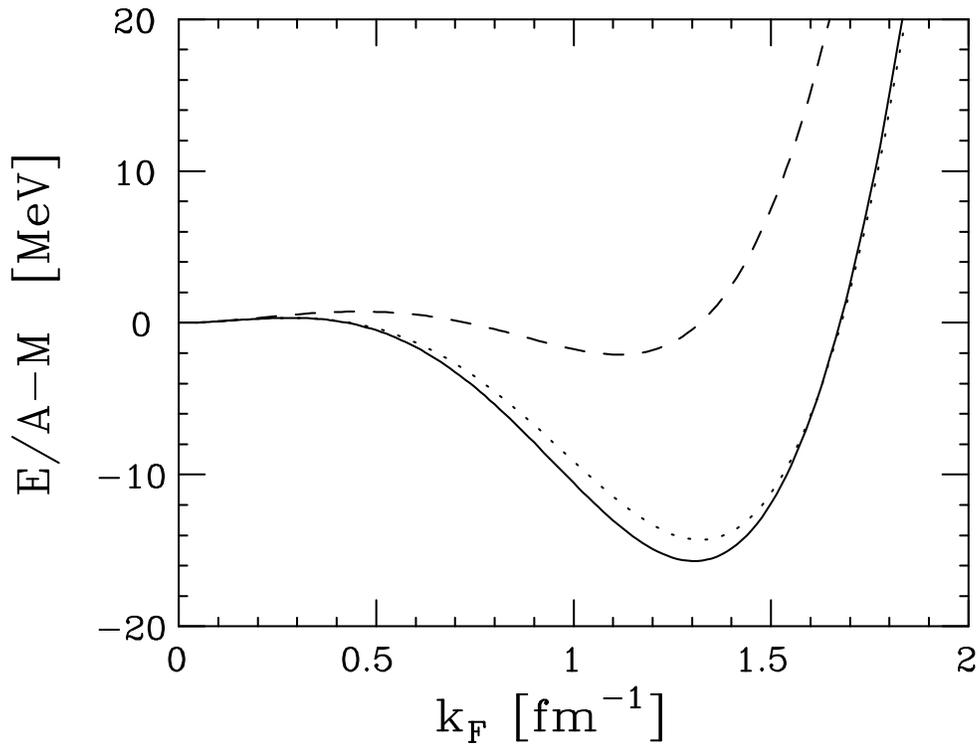}}
\vspace{-3.0cm}
\caption{The energy per particle including the $\sigma$, $\omega$ and 
$\pi$ mesons. The solid curve is the Hartree result, and the dashed (dotted) 
curve is the HARTREE+FOCK result (contact interaction subtracted). }
\end{figure}

\begin{figure}
\epsfxsize=15.cm
\epsfxsize=15.cm
\centerline{\epsfbox{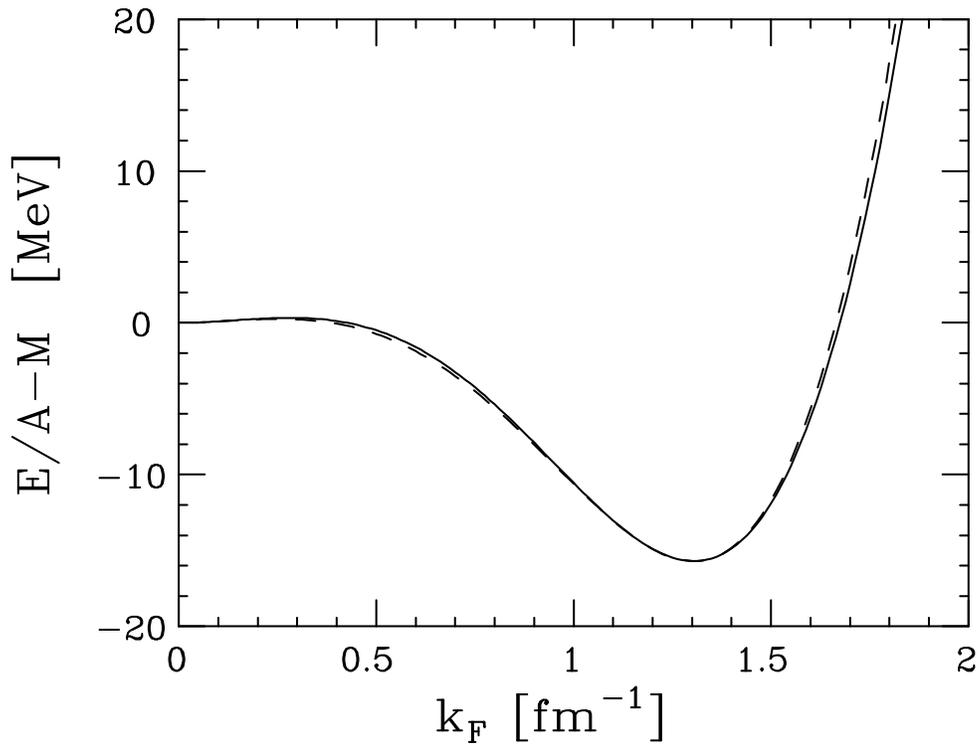}}
\vspace{-3.0cm}
\caption{The energy per particle including the $\sigma$, $\omega$, $\pi$ and 
$\rho$ mesons. The solid curve is the HARTREE result, and the dashed 
curve is the HARTREE+FOCK result for the parameter set (b) on Table~I. Note
that the coupling constants for the HARTREE+FOCK result have been readjusted
to produce the same saturation density and energy per nucleon.}
\end{figure}

\begin{figure}
\epsfxsize=15.cm
\epsfxsize=15.cm
\centerline{\epsfbox{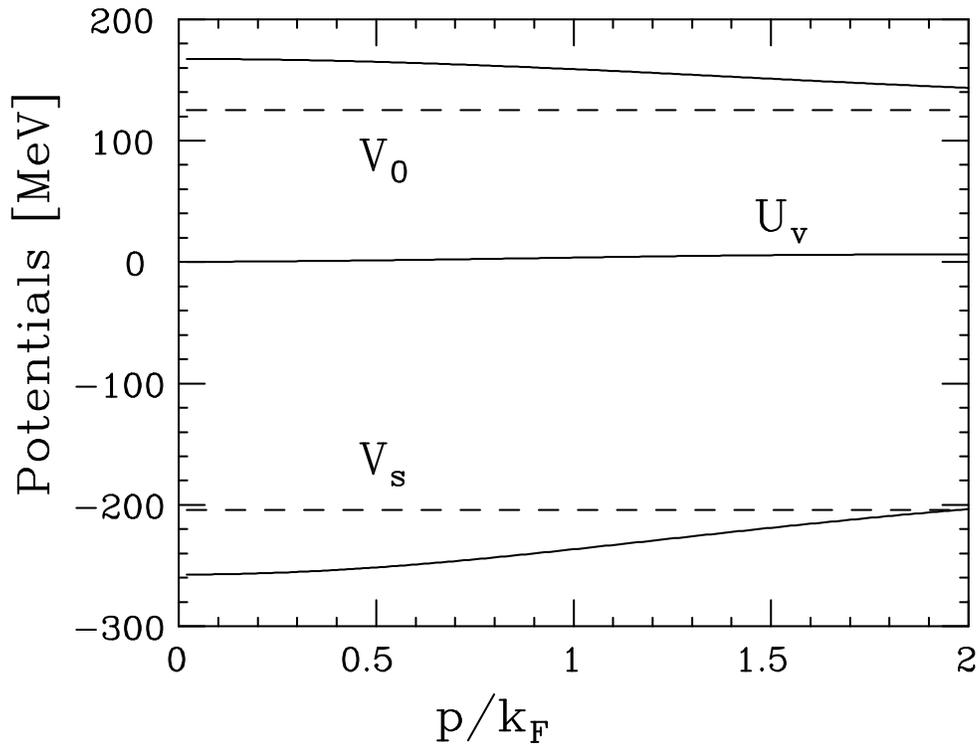}}
\vspace{-3.0cm}
\caption{Single particle potentials $V_s=-g_{\sigma}\,\sigma_0+U_s$, 
$V_0=g_{\omega}\,\omega_0+U_0$, and $U_v$ as a function of $p/k_F$ at the
saturation density, $\rho_0=0.15$fm$^{-3}$. The solid lines are the 
HARTREE+FOCK results for parameter set (b) in Table~I, and the dashed 
lines are the HARTREE results.}
\end{figure}

\end{document}